\documentclass[12pt,preprint]{aastex}

\slugcomment{Submitted to the Astrophysical Journal}

\lefthead{S.S. Hasan, A. A. Van Ballegooijen, W. Kalkofen \& O. Steiner }
\righthead{Dynamics of the Magnetic Network: Two-dimensional MHD Simulations}
\shorttitle{Dynamics of the Magnetic Network}
\shortauthors{Hasan et al.}

\begin{document}


\title{Dynamics of the Solar Magnetic Network\\ Two-dimensional MHD Simulations }

\author{S. S. Hasan} 
\affil{Indian Institute of Astrophysics, Bangalore-560034, India (hasan@iiap.res.in)}

\author{A. A.  van Ballegooijen and W. Kalkofen}
\affil{Harvard-Smithsonian Center for Astrophysics, 60 Garden Street,
Cambridge, MA~02138, U.S.A.}
\and
\author{O. Steiner, Kiepenheuer Institut f\"ur Sonnenphysik,
7800 Freiburg, Germany}

\begin{abstract}
The aim of this work is to identify the physical processes that occur
in the network and contribute to its dynamics and heating.  We model
the network as consisting of individual flux tubes with a
non-potential field structure that are located in intergranular lanes.
With a typical horizontal size of 200 km at the base of the
photosphere, they expand upward and merge with their neighbors at a
height of about 600 km.  Above a height of approximately 1000 km the
magnetic field starts to become uniform. 

Waves are generated in this medium by means of  motions at the lower
boundary. We focus on transverse driving, which generates both fast
and slow waves within a flux tube and acoustic waves at the interface
of the tube and the field-free medium.  The acoustic waves at the
interface are due to compression of the gas on one side of the flux
tube and expansion on the other. These waves travel upward along the
two sides of the (2D) flux tube and enter it, where they become
longitudinal waves. For impulsive excitation with a time constant of
120~s, we find that a dominant feature is the creation of vortical
motions that propagate upward. We have identified an efficient
mechanism for the generation of longitudinal waves and shock formation
in the chromospheric part of the flux concentration. We examine some
broad implications of our results.

\end{abstract}

\section{Introduction}
The quiet solar chromosphere is bifurcated into the magnetic network
on the boundary of supergranulation cells and the largely field-free
internetwork medium in the cell interior, with respective filling
factors in the ratio of 2 to 3 in the middle chromosphere (i.e., at a
height of about 1~Mm above the level of unit optical depth at 5,000
{\AA}). The network is about 30\% brighter in Ca~II K$_{\rm 2v}$
emission than the internet chromosphere (Skumanich, Smythe \& Frazier
1975). Both the network and the internet medium show {\it bright
points} (BPs), which are prominent in the emission peaks in the cores
of the Ca~II H and K lines, formed in the middle chromosphere.
However, the dynamical and spectral properties of network and internet
BPs are quite different. In internetwork areas the chromospheric
velocity power spectrum is dominated by oscillations with frequencies
at and above the acoustic cutoff frequency (period 3 min in upper
photosphere, e.g., Rutten \& Uitenbroek 1991), whereas in the network,
Ca II H line center velocity and intensity power spectra are dominated
by low-frequency oscillations with periods of 7-20~min (Lites, Rutten
\& Kalkofen 1993). These long-period waves have also been observed at
larger heights (Curdt \& Heinzel 1998; McAteer et al.\ 2002, 2003).
Furthermore, network BPs show high emission in the line core most of
the time (Lites et al.\ 1993), whereas internetnetwork locations 
have a bright
phase only about 5\% to 10 \% of the time (von Uexk\"ull \&
Kneer 1995).  Finally, 80\% of network BPs have symmetric profiles in
the line core, with more or less equal intensity in the blue and red
emission peaks on either side of the central absorption. In contrast,
only about 30\% of internet BPs have symmetric profiles, and 40\% have
blue-peak enhancements (Grossmann-Doerth, Kneer \& v.Uexk\"ull 1974);
hence the name H$_{\rm 2v}$ or K$_{\rm 2v}$ bright point.

While the qualitative properties of internetwork BPs are reasonably well
understood, including their formation in upward-propagating acoustic shocks
that encounter down\-ward-flowing gas (Carlsson \& Stein 1995), this is not
true for network BPs. The physical processes that heat the magnetic network
have not been fully identified. Are network BPs heated by wave dissipation, and
if so, what is the nature of these waves? How can we understand the observed
relatively constant emission and symmetric line profiles of network BPs? The
source of energy for network BPs is likely to be magnetohydrodynamic waves.
Possible candidates are: 
{\singlespace 
\begin{itemize} 
\item Kink (transverse) waves, generated inside flux tubes by the
buffeting action of granules; 
\item Longitudinal waves, generated by pressure fluctuations inside
flux tubes; 
\item Torsional (Alfv\'en) waves, generated inside flux tubes; 
\item  acoustic waves, generated in the field-free atmosphere
surrounding flux tubes, that penetrate into the tubes; 
\item Acoustic waves, generated at the interface of flux tubes and the
outside medium, that also penetrate into the flux tubes.  
\end{itemize}
} 
The first three are well-known flux tube modes (e.g., Spruit 1981).
Several investigations have focused on the generation and propagation
of transverse and longitudinal wave modes and their dissipation in the
chromosphere (e.g., Zhugzhda, Bromm \& Ulmschneider 1995; Fawzy et
al.\ 2002; Ulmschneider 2003 and references therein).  Torsional waves
have received some attention (e.g., Hollweg 1982; Noble, Musielak \&
Ulmschneider 2003).  The fourth wave type is an obvious source since
it is responsible for heating the internetwork medium.  However, the
importance of the fifth wave type, which we found during the course of
this investigation, does not appear to have been adequately recognized
earlier.  It will be discussed in more detail later on.

The present study is a continuation of earlier work on the excitation
of transverse and longitudinal waves in magnetic flux tubes by the
impact of fast granules on flux tubes (Hasan \& Kalkofen 1999), as
observed by Muller \& Roudier (1992) and Muller et al.\ (1994), and
following the investigation by Choudhuri, Auffret \& Priest (1993),
who studied the generation of kink waves by footpoint motion of flux
tubes. The observational signature of the modeled process was highly
intermittent in radiation emerging in the H and K lines, contrary to
observations. By adding waves that were generated by high-frequency
motion due to the turbulence of the medium surrounding flux tubes
(Hasan, Kalkofen \& van Ballegooijen 2000), the energy injection into
the gas inside a flux tube became less intermittent, and the time
variation of the emergent radiation was in better agreement with the
more steady observed intensity from the magnetic network.

The above studies modeled wave excitation and propagation in terms of
the Klein-Gordon equation, motivated by the identification of the
power peak near 7 min in the observed power spectrum (Lites et al.\
1993; Curdt \& Heinzel 1998; McAteer et al.\ 2002) with the cutoff
period of kink waves in thin magnetic flux tubes (Kalkofen 1997).
This analysis was based on a linear approximation in which the
longitudinal and transverse waves are decoupled.  However, it is well
known that the velocity amplitude $v(z)$ for the two modes increases
with height $z$ (for an isothermal atmosphere, as $v \propto\exp
[z/4H]$, where $H$ is the pressure scale height), so that the motions
are expected to become supersonic higher up in the atmosphere. At such
heights, nonlinear effects become important, leading to a coupling
between the transverse and longitudinal modes.  Some progress on this
question has been made in one dimension, using the nonlinear equations
for a thin flux tube, by Ulmschneider, Z\"ahringer \& Musielak (1991),
Huang, Musielak \& Ulmschneider (1995), and Zhugzhda, Bromm \&
Ulmschneider (1995), and more recently by Hasan et al.\ (2003) and
Hasan \& Ulmschneider (2004), who examined mode coupling between
transverse and longitudinal modes in the magnetic network. By solving
the nonlinear, time-dependent MHD equations it was found that
significant longitudinal wave generation occurs in the photosphere,
typically for Mach numbers as low as 0.2, and that the onset of shock
formation occurs at heights of about 600 km above the photospheric
base, accompanied by heating (Hasan et al. 2003, Hasan \& Ulmschneider
2004).  The efficiency of mode coupling was found to depend on the
magnetic field strength in the network and is a maximum for field
strengths corresponding to $\beta\approx 0.2$, when the kink and tube
wave speeds are almost identical. This can have interesting
observational implications. Furthermore, even when the two speeds are
different, once shock formation occurs, the longitudinal and
transverse shocks exhibit strong mode coupling.

The above studies on the magnetic network make use of two important
idealizations: they assume that the magnetic flux tubes are thin, an
approximation that becomes invalid at about the height of formation of
the emission peaks in the cores of the H and K lines; and they neglect
the interaction of neighboring flux tubes.  Some progress in this
direction has recently been made by Rosenthal et al. (2002) and Bogdan
et al. (2003), who studied wave propagation in a two-dimensional
stratified atmosphere, assuming a potential magnetic field to model
the network and internetwork regions on the Sun. They examined the
propagation of waves that are excited from a spatially localized
source in the photosphere. Their results indicate that there is strong
mode coupling between fast and slow waves at the so-called magnetic
canopy, which they identify with regions where the magnetic and gas
pressures are comparable. As a consequence of the potential-field
approximation, some magnetic field lines are nearly horizontal even at
the base of the field. Such a model may not be appropriate for a
network patch, which is perhaps better idealized as a collection of
vertical tubes (Cranmer \& van Ballegooijen 2005).


Thus the problem that we address concerns wave propagation in regions
that are largely representative of individual structures in the
magnetic network --- this is different from the one analysed by
Rosenthal et al.\ and Bogdan et al. Our initial configuration consists
of flux tubes in 2-D magnetostatic equilibrium with a sharp interface
between the tube and the surrounding gas. Waves are generated in this
medium by means of transverse motion at the lower boundary, which
displaces the entire flux tube, unlike the problem studied in the
above papers where the source region is confined to a portion of the
magnetic structure. This can have interesting consequences, some of
which were unrecognized thus far. These will be explored in later
sections of the paper. Our calculations indicate the presence of a new
and efficient mechanism for longitudinal wave generation and shock
formation in the chromosphere.

The present study forms the first of a series devoted to a detailed
investigation of wave propagation in the magnetic network. As a first
step, we go beyond the thin flux tube approximation and employ a
two-dimensional treatment in slab geometry, similar to Rosenthal et
al.\ and Bogdan et al.\ It is not possible (within the 2-D framework) to
examine kink and sausage waves, which would require a
three-dimensional  treatment, but, nevertheless, there are still many
important physical effects that this approach allows us to examine.
For simplicity, we neglect non-adiabatic effects at this stage, which
we will include in a separate paper.

Several papers have looked at MHD waves in various geometries in the
solar atmosphere using multi-dimensional simulations (e.g., Shibata
1983; Cargill, Spicer, \& Zalesak 1997; Ofman \& Davila 1998; Sakai,
Igrarashi \& Kawata 1998; Ofman, Nakariakov, \& DeForest 1999;
additional references can be found in Bogdan et al. 2003).

The organization of this paper is as follows: in Sect.\ 2 we describe
the initial equilibrium model, its construction and in Sect.\ 3 the
numerical method of solution along with the driving mechanism. The
results of our calculation are presented in Sect.\ 4 followed by a
discussion and summary in Sect.\ 5.

\section{Model}
\begin{figure} 
\epsscale{0.8}
\plotone{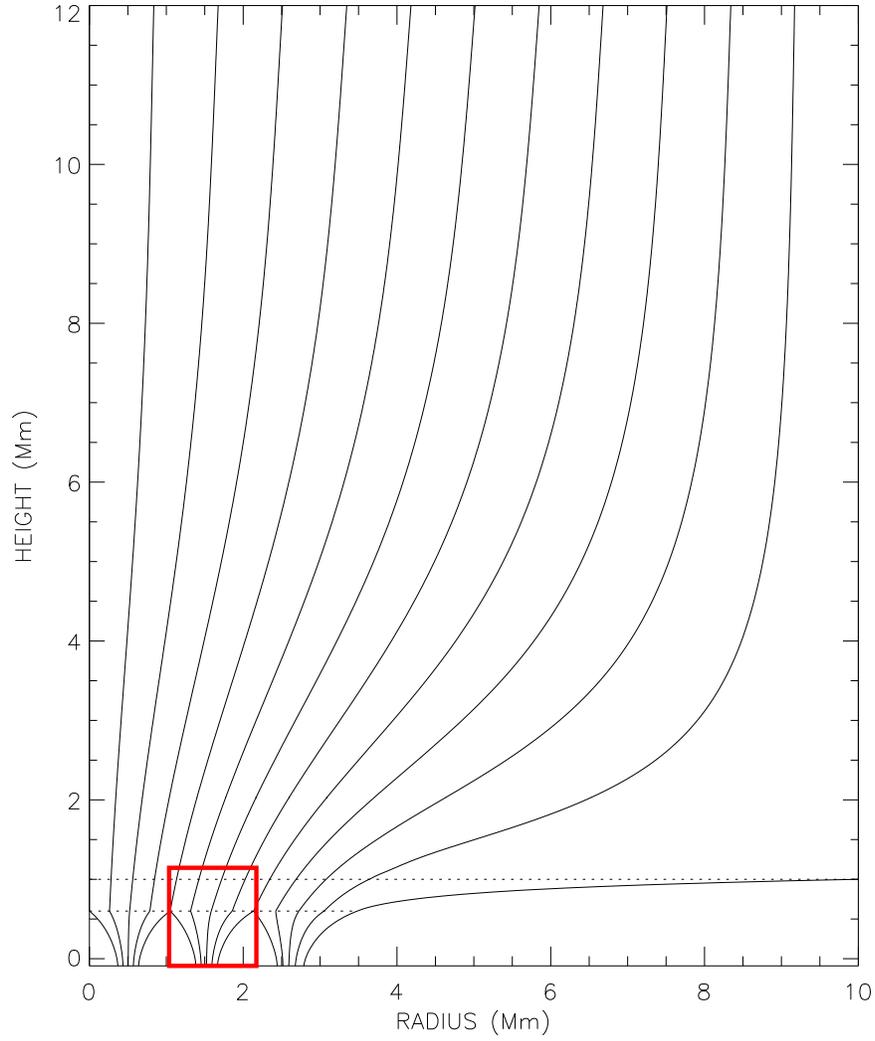}
\caption{
Model of a network element consisting of individual flux tubes
separated at the photospheric surface by a distance of 1000 km which
merge at a height of about 600 km.  The red box corresponds to the
domain taken up for dynamical simulations.  
\label{fig1}} 
\end{figure}

Following Cranmer \& van Ballegooijen (2005),  we treat 
a network element (typical flux $\sim 3 \times 10^{19}$ Mx)
to consist of a collection of smaller flux tubes that are spatially
separated from one another in the photosphere. The gas pressure in the
atmosphere decreases with increasing height, causing a lateral
expansion of the flux tubes.  Neighboring flux tubes within the
network element merge into a monolithic structure at some
height. Above this ``merging height'' the network element consists of
a single thick flux tube that further expands with height. The outer
edge of this tube forms a magnetic canopy that overlies the
neighboring supergranular cells. A second merging occurs when
neighboring network elements come together at a canopy height.
Figure \ref{fig1} schematically shows the picture we have for the network field
structure. It consists of three distinct regions:
\begin{enumerate}
\item Photospheric region up to about 0.6 Mm, consisting of individual
flux tubes, with a typical diameter of 200 km in the low photosphere.
Their foot points are located  in intergranular lanes and separated
from one another by about the diameter of a granule ($\sim 1$ Mm).
They expand upward and merge with their neighbors at a height of
about 0.6 Mm.
\item Lower chromosphere, between the heights of 0.6 and 1 Mm, the
merged network flux element expands laterally over the surrounding
supergranular cell-center and overlying field-free chromosphere;
\item Upper chromosphere and corona, between 1 and 12 Mm, the fully
merged magnetic field fills the available volume. At larger heights,
the field expands primarily in the vertical direction and becomes
more or less uniform. However, at lower heights, between 1 and 2 Mm,
the field strength varies significantly with horizontal position,
and the field strength directly above the flux tubes (left side of
Figure \ref{fig1}) is much larger than above the supergranular cell
center (right side of Figure \ref{fig1}).
\end{enumerate}

Our model is based on the idea that the base of flux tubes is located
in subsurface layers where convective flows may be different from
those in the visible photosphere. Flux tubes occur in regions with
convective downflows below intergranular lanes. These downflows are
likely to be highly turbulent, involving lateral motions that
produce transverse waves in the flux tubes. When the upward
propagating waves reach the photosphere, they cause horizontal motions
of flux tubes relative to their local surroundings. This generates
excess pressure on the leading edge of a flux tube, and a pressure
deficit on the trailing edge. Our  two-dimensional MHD calculations
(see section Sect.\ 3) indicate that these pressure pulses produce an
upward-velocity pulse on the leading side of the flux tube and a
downward pulse on the trailing side. These pressure pulses and
vertical flows are an integral part of the MHD wave, so the transverse
and longitudinal motions are strongly coupled.  

\subsection{Initial Two-dimensional Magnetostatic Model} \label{equilibrium}
Let us consider an individual flux tube at the base of a magnetic
network element on the quiet sun; the region of interest is indicated
by the small box in Figure \ref{fig1}.  At heights below about 600 km,
the flux tubes are spatially distinct from one another, and are
embedded in a field-free ``external'' medium.  At $\approx 1000$ km,
the flux tubes merge into a more uniform field (upper part of small
box). The structure of the flux tube at the initial instant is assumed
to be in static equilibrium and is determined by the magnetostatic
force balance equation:

\begin{equation}
- \nabla p + \rho {\bf g} + \frac{1}{4\pi} ( \nabla \times {\bf B} )
\times {\bf B} = 0 , \label{eq:force}
\end{equation}
where ${\bf g} = -g \hat{\bf z}$ is the gravitational acceleration,
$p$ is the gas pressure, $\rho$ is the density, and ${\bf B}$ is the
magnetic field. The third term in (\ref{eq:force}) describes the
Lorentz force due to electric currents at the boundary between the
flux tube and its local surroundings.

We have developed a numerical code for solving equation (\ref{eq:force}) in
two dimensions; all quantities are assumed to be independent of the
horizontal $y$ coordinate, so the flux tubes are approximated by
sheets. We consider a rectangular domain $x=[0,L]$ and $z=[0,H]$,
representing one half of a flux sheet; $x=0$ is the flux sheet axis,
$x=L$ is the interface with the neighboring sheet on the right, and
$z=0$ is the base of the photosphere. The magnetic field is written
in terms of a flux function $A(x,z)$:
\begin{equation}
B_x = - \frac{\partial A} {\partial z} , ~~~~
B_z = \frac{\partial A} {\partial x} .  \label{eq:mag}
\end{equation}
The boundary conditions are $A(0,z)=0$ and $A(L,z)=A_{\rm max}$, where
$A_{\rm max}$ is one half of the total magnetic flux within the sheet.
The gas pressure and density are written as:
\begin{eqnarray}
p & = & p_{\rm int}(z) \left[ 1 + \beta_0^{-1} F(A) \right] ,
\label{eq:pres} \\
\rho & = & \rho_{\rm int}(z) \left[ 1 + \beta_0^{-1} F(A) \right] ,
\label{eq:dens}
\end{eqnarray}
where $p_{\rm int} (z)$ is the internal gas pressure as function of height
along the axis of the flux sheet; $\rho_{\rm int}(z)$ is the internal
density, $\rho_{\rm int} (z) = - g^{-1} dp_{\rm int}/dz$; $\beta_0$ is the
ratio of gas- and magnetic pressures at the base (on axis); and $F(A)$
is a function describing the variation of gas pressure across field
lines. Note that the temperature depends only on height ($p/\rho$ is
independent of $x$). The function $F(A)$ varies from zero on the axis
of the sheet to $F(A_{\rm max})=1$ in the external medium, resulting in
distributed electric currents at the interface between the sheet and
its local surroundings. 

Inserting (\ref{eq:mag}), (\ref{eq:pres})
and (\ref{eq:dens}) into (\ref{eq:force}) yields
\begin{equation}
\nabla^2 A + 4 \pi p_{\rm int}(z) \beta_0^{-1} \frac{dF}{dA} = 0 ,
\label{eq:flux}
\end{equation}
which can be solved by minimizing the following Lagrangian:
\begin{equation}
W = \int_0^H \int_0^L \left[ \frac{1}{2} | \nabla A |^2 - 4 \pi
p_{\rm int}(z) \beta_0^{-1} F(A) \right] ~ dx ~ dz . \label{eq:func}
\end{equation}

The minimization is done by varying the $A$-values on a grid of 120 by
240 cells, using the conjugate-gradient method (Press et al.\ 1992). A
similar technique was used in (Hasan et al. 2003), where we
constructed a model of a very thin flux tube (see the Appendix of this
paper), and in Cranmer \& van Ballegooijen (2005). In these papers a
Lagrangian description was used, and the radial positions $r(\Phi,z)$
varied on a fixed grid of $\Phi$ (the flux function in cylindrical
coordinates) and height $z$. In the present paper we vary $A(x,z)$ on
a fixed grid of $x$ and $z$. 

A model containing one whole flux sheet can be obtained by mirroring
the field with respect to $x=0$, and models containing multiple sheets
can be obtained by further mirroring with respect to $x=L$. The
resulting fields will be used as initial conditions for the
two-dimensional MHD calculation (see next subsection).

We perform calculations for a single sheet using the
following form for $F(A)$:
\begin{equation}
F(A) = \left\{ \begin{array}{ll}
\frac{4}{3} \left( A/A_{\rm max} \right)^2 - \frac{1}{3}
\left( A/A_{\rm max} \right)^8 & \mbox{if $A \leq A_{\rm max}$} , \\
1 & \mbox{otherwise.} \end{array} \right.
\end{equation}
This produced a smooth transition of gas pressure from the interior to
the exterior of the sheet, and the residual magnetic field in the
external medium was very small ($\sim 1$ Gauss). The internal pressure
as function of height was approximated as a sum of two exponentials:
\begin{equation}
p_{\rm int} (z) = p_1 \exp(-z/H_1) + p_2 \exp(-z/H_2) ,
\end{equation}
with a photospheric pressure scale height of $H_1$ = 110 km and a
chromospheric scale height of $H_2$ = 220 km. To obtain kilogauss fields
in the photosphere, we used $\beta_0 = 0.5$, so that the external gas
pressure is three times the internal gas pressure. 

In the present calculations we consider a full flux sheet which is
placed symmetrically with respect to the midpoint of the computational
domain, which we take to be a square of size 1200 km with a uniform
grid of 240 $\times$ 240 cells, corresponding to a mesh spacing of
5~km in either direction.  The axis of the flux tube is now located at
$x=600$ km.  The width of the sheet at $z=0$ is approximately 200 km,
and the central field strength is 1530 G. 

\begin{figure} 
\plotone{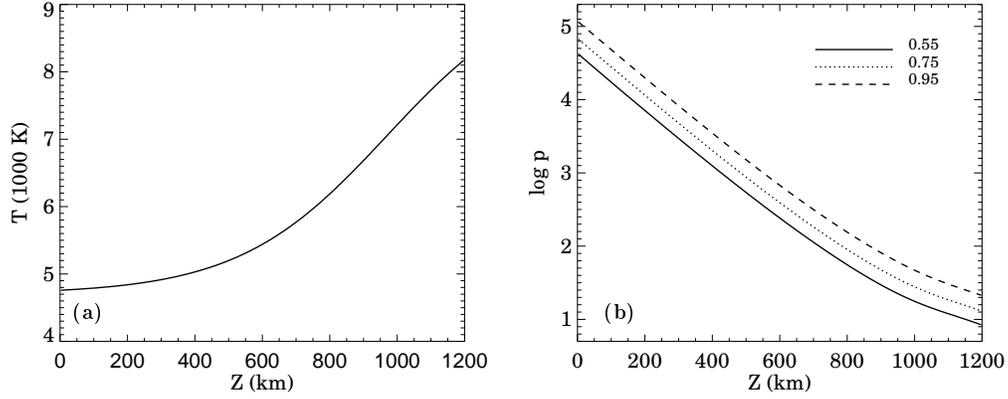}
\caption{
Variation of (a) the temperature $T$,  and (b) the logarithm of the
pressure $p$ as a function of height on the field lines corresponding
to $f=0.55$ (solid curve), 0.75 (dotted curve) and 0.95 (dashed curve)
in the equilibrium configuration. 
Note the temperature is uniform (by assumption) in the horizontal direction. 
\label{fig2}}  
\end{figure}

Table 1 provides some of the parameters of the
equilibrium model on the tube axis and in the ambient medium at the
base ($z=0$) and the top ($z=1200$ km) of the computational domain.
Figures \ref{fig2}(a) and (b) depict the variation with height of the
temperature and pressure, respectively.  It should be noted that the
temperature by assumption is constant in the horizontal direction.
The sound speed $c_S$ corresponding to the temperature shown in Figure
\ref{fig2}(a) varies from 7.2 km s$^{-1}$ at $z=0$ to 9.3 km s$^{-1}$
at $z=1200$~km. In Figure \ref{fig2}(b), the pressure is shown on
different field lines corresponding to a fractional flux $f \equiv
A/A_{\rm max}$ as measured from the left edge of the computational
domain.  Thus, $f=0.5$ corresponds to the tube axis.  As expected, the
gas pressure on each field line increases as one approaches the tube
boundary.  The dashed curve ($f=0.95$) essentially depicts the height
variation of the pressure in the ambient medium.

\begin{figure} 
\plotone{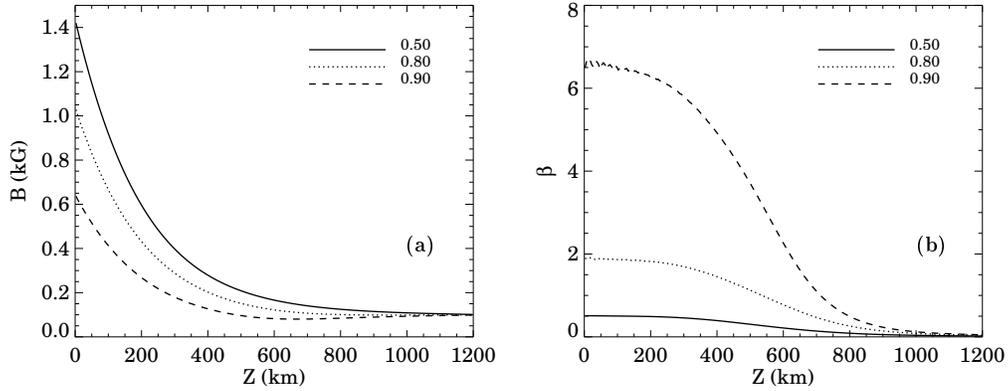}
\caption{
Variation of (a) the magnetic field strength $B$,  and (b) $\beta$
as a function of height on the field lines corresponding
to $f=0.5$ (solid curve), 0.8 (dotted curve) and 0.9 (dashed curve)
in the equilibrium configuration. 
\label{fig3}}
\end{figure}

Figures \ref{fig3}(a) and (b) show the height variation of the
magnetic field strength $B$ and of $\beta$ on various field lines,
parameterized, by $f=0.5,0.8,0.9$.  The field strength on each of the
field lines drops rapidly with $z$ in the first few hundred kilometers
after which it approaches a constant value of about 100 G.  On the
other hand, $\beta$ is practically constant with $z$ in the lower
region of the atmosphere, which is similar to the behavior one finds
in an thin flux tube with equal internal and external temperature at
the same height (e.g. see Hasan et al. 2003), and where both $B^2$ and
$p$ have the same height dependence. In the upper part of the flux
tube, $B$ is constant and $\beta$ drops off sharply with $z$,
essentially mimicking the $p$ dependence. The Alfv\'en speed, which is
related to $\beta$ by the approximate relation $v_A =
c_S/\sqrt{\beta}$, is almost constant (with height) in the lower
regions of the atmosphere in the tube, but increases sharply with $z$
in the higher layers.  

\begin{figure} 
\epsscale{0.8}
\plotone{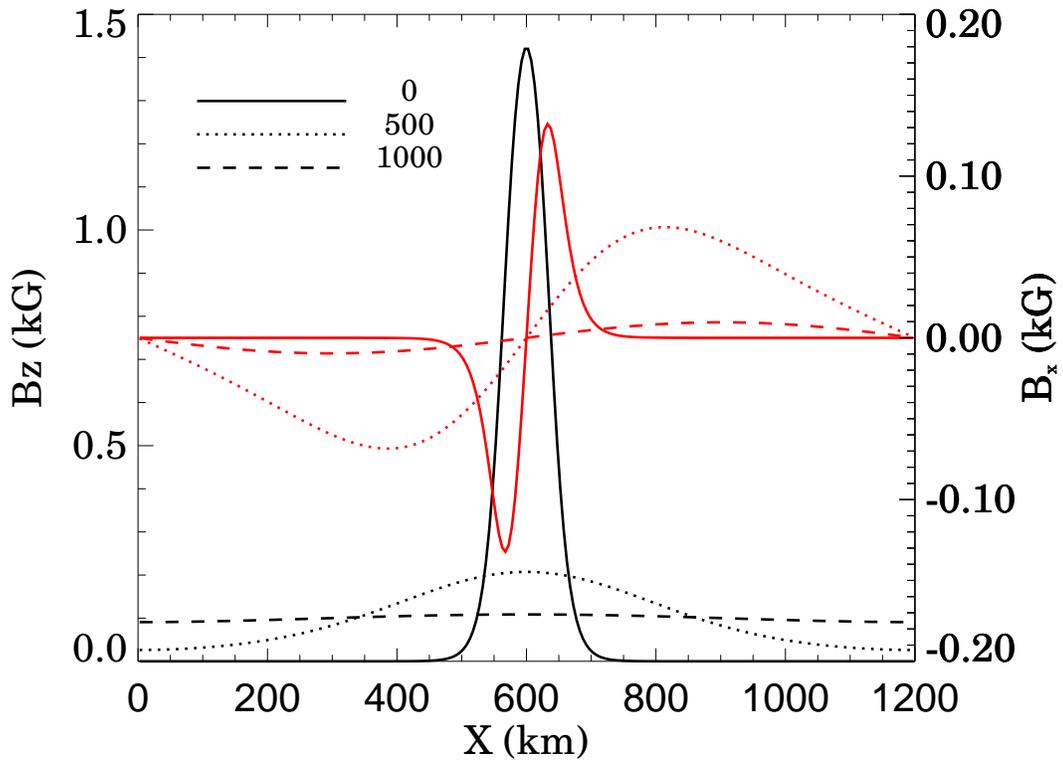} 
\caption{
Variation of $B_z$ with horizontal distance (from the left edge of the
computational domain) at the following heights: $z=$0 (black solid
line), 500 km (black dotted line) and 1000 km (black dashed line). The
corresponding red curves denote $B_x$ measured with respect to the
values given on the right scale.  \label{fig3n}} \end{figure}

Figure \ref{fig3n} shows the variation with horizontal distance $x$ of
the vertical component of the magnetic field $B_z$ at various heights
$z=$0 (black solid line), 500 km (black dotted line) and 1000 km
(black dashed line). The red curves show the corresponding variation
of $B_x$ measured with respect to the scale on the right. The solid
lines (black and red) clearly show that the magnetic field at the base is
confined to a width of 200 km. At a height of $1000$~km, $B_x$ becomes
very small (a few Gauss) and the field is essentially vertical and
constant (approximately 100 G) in the horizontal direction.

\section{Method and Boundary Conditions} We consider wave generation
in the configuration described in the previous section by perturbing
the lower boundary and solving the two-dimensional magnetohydrodynamic
(MHD) equations in conservation form for an inviscid adiabatic fluid.
These consist of the usual continuity, momentum, entropy (without
sources) and magnetic induction equations (for details see Steiner,
Kn\"olker \& Sch\"ussler 1994). The unknown variables are the density,
momentum, entropy per unit mass and the magnetic field.  We assume
that the plasma consists of fully ionized hydrogen with a mean
molecular weight of 1.297. The temperature is computed from the
specific entropy and the pressure is found using the ideal gas law.   

The above MHD equations are solved following the numerical procedure
given by Steiner et al. (1994).  Briefly, the equations are
discretized on a two-dimensional mesh using a finite-volume method
which has the advantage of preserving div~${\bf B}=0$ to machine
accuracy. The method employs finite differences to compute numerical
fluxes based on the flux-corrected transport (FCT) scheme of Oran and
Boris (1987).  The time integration is explicit and has second-order
accuracy in the time step. Small time steps are required to satisfy
the Courant condition in the upper part of the domain, where the
Alfv\'en speed is large. 

Periodic boundary conditions are used at the horizontal boundaries.
At the top boundary, (a) the vertical and horizontal components of
momentum are set to zero; (b) the density is determined using linear
logarithmic extrapolation; (c) the horizontal component of the
magnetic field and temperature are set equal to the corresponding
values at the preceding interior point. The vertical component of the
magnetic field is determined using the condition div $B$=0. Similar
conditions are used at the lower boundary, except for the density,
temperature and horizontal component of the velocity (or momentum).
The density is obtained using cubic spline extrapolation, the
temperature is kept constant at its initial value and $v_x$ at $z=0$
is specified as follows:
\begin{equation}
v_x(x,0,t)=v_0\sin (2\pi t/P),
\label{eq:a1}
\end{equation}
where $v_0$ denotes the amplitude of the horizontal motion and $P$ the
wave period. This form was chosen to simulate the effect of 
transverse motion of the lower boundary. For simplicity, we assume
that all points at the lower boundary have this motion, since this
does not generate any waves in the ambient medium, other than at
the interface with the flux tube, as we shall see later on.

\section{Dynamics of a flux sheet} 
In Sect.\ 2, we presented a model for a single flux sheet in static
equilibrium.  The stability of this equilibrium was checked by solving
the time-dependent two-dimensional MHD equations without any external driving
(assuming rigid boundaries in the vertical direction). We found that
this equilibrium was maintained to high accuracy. The maximum
amplitude of the flows was no more than a few meters per second over
time scales greater than the sound travel time (in both the vertical
and horizontal directions), corresponding to over 10000 time steps. 

Let us now consider wave generation in the equilibrium atmosphere by
means of a transverse motion at the lower boundary ($z=0$) of the
atmosphere, which displaces the flux sheet. We focus on two limiting
cases, corresponding to impulsive and periodic excitation,
respectively.  

\subsection{Impulsive excitation}
\begin{figure} 
\epsscale{1.2}\hspace*{-0.7in}
\plotone{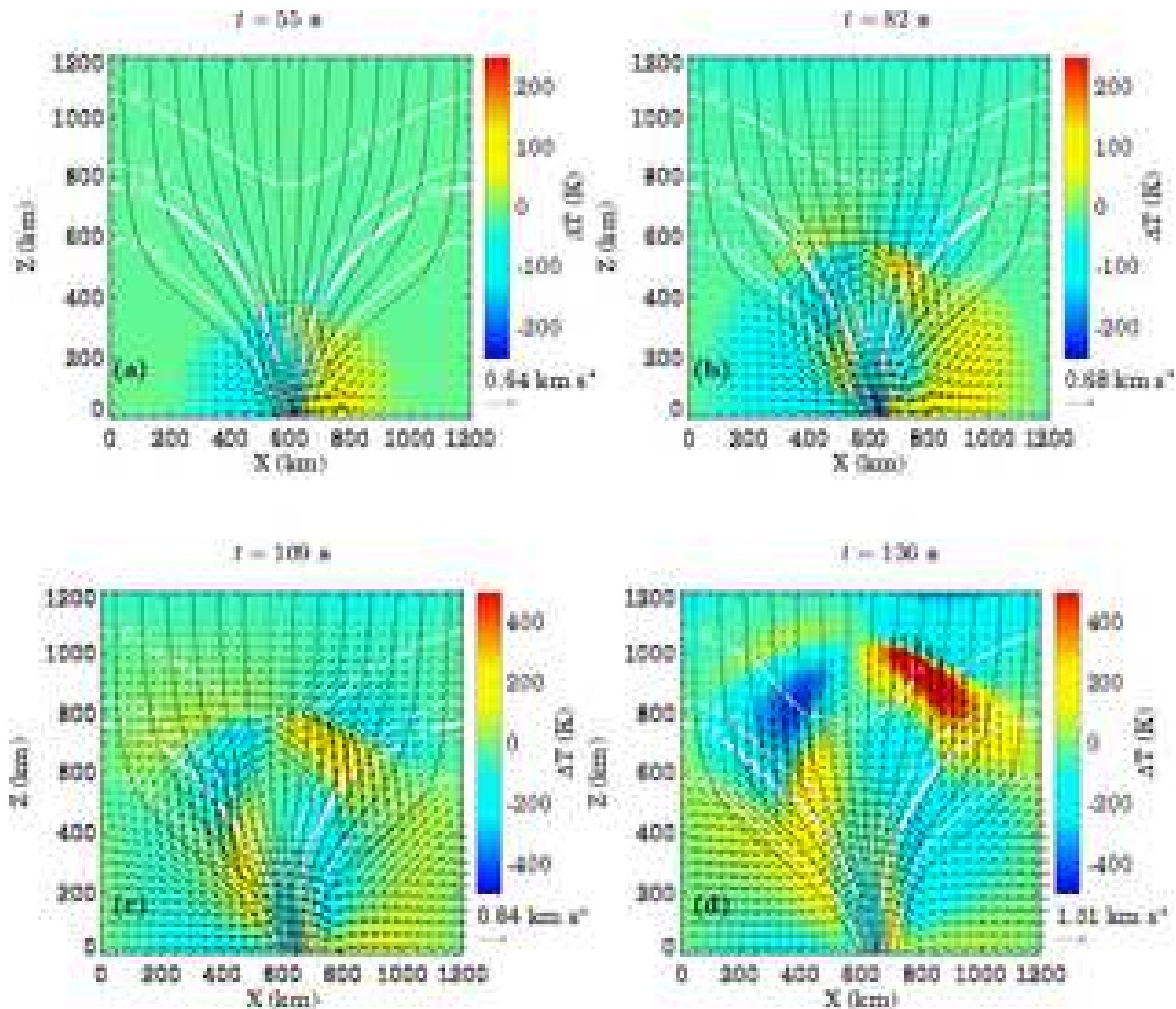}  
\caption
{Flow pattern  and temperature perturbation, , $\Delta T$,  (about the initial
state) at (a) 55~s, (b) 82~s, (c) 109~s and (d) 136~s
in a network element due to  horizontal motion at the lower boundary,
with an amplitude of 750 m s$^{-1}$, applied for half a wave period
($P=240$~s), after which the motion ceases.  The thin black curves are
the magnetic field lines, the arrows denote the direction of the flow,
and the color scale shows the temperature perturbation.  The white
curves denote contours of constant $\beta$ corresponding to
$\beta=0.1$, 0.5, 1.0 (thick curve) and 10.   
\label{fig4}} 
\end{figure}

This case corresponds to uniform displacement of the lower boundary
at $z=0$ to the right which lasts half a wave period ($P = 240$~s) and
then stops. The peak transverse velocity is $v_0 = 750$~m s$^{-1}$.
Figures \ref{fig4}(a)-(d) show, at times of 52~s, 82~s, 109~s and 136~s,
respectively, the velocity field (arrows), the magnetic field (black
lines) and the temperature change $\Delta T=T-T_0$, where $T_0$ is the
initial temperature at each height. The maximum value of the
velocity and the color table for the temperature are shown on the right
of each figure.  We have omitted velocities smaller than 30 m
s$^{-1}$.  The white lines denote contours of constant beta for 
values of $\beta=$ 0.1, 0.5, 1.0 (thick line) and 10. 

The horizontal motion of the flux tube at the lower boundary pushes
the field lines uniformly to the right.  In the flux tube, this
motion is communicated to the upper layers as a fast mode, which
travels with the local Alfv\'en speed.  The transverse motion of the
field lines compresses the gas on the right interface of the tube
and the ambient medium (above the base, where there is no horizontal
motion in the field free medium).     

\begin{figure} 
\epsscale{0.8}
\plotone{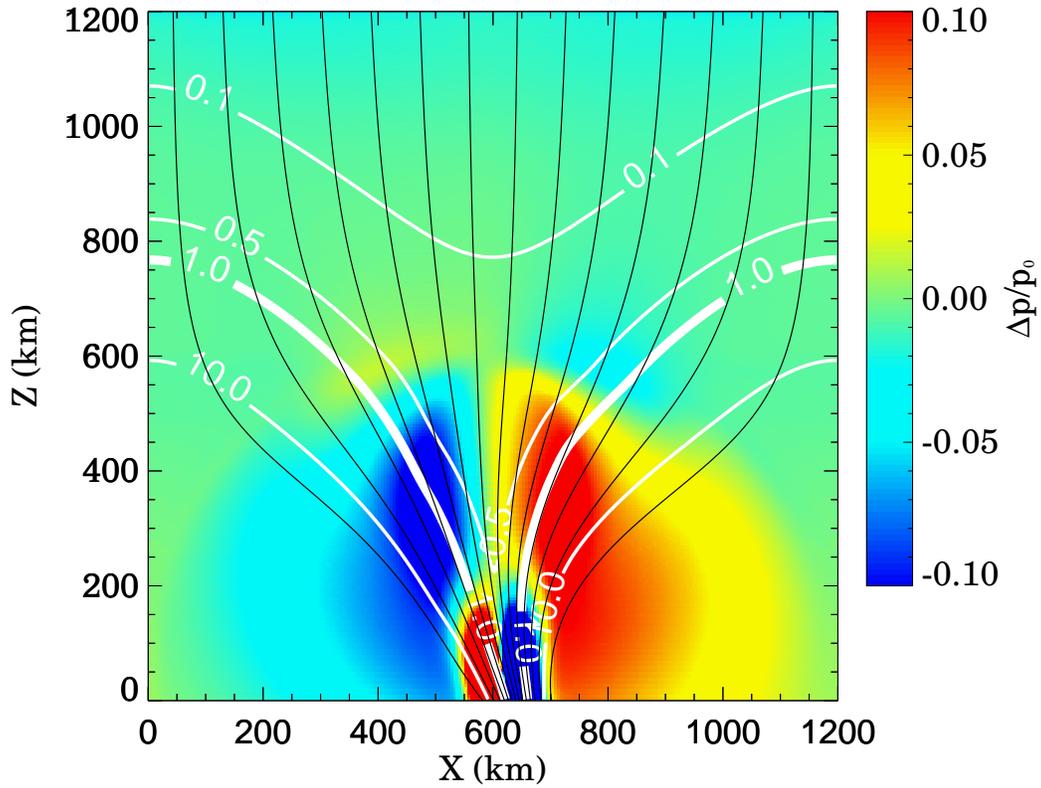} 
\caption
{The relative pressure perturbation (about the initial
state), $\Delta p/p_0$, at $t=82$~s
in a network element due to horizontal motion at the lower boundary
with an amplitude of 750 m s$^{-1}$ and a period of $P=240$~s.
The thin black curves are
the magnetic field lines and the white
curves denote contours of constant $\beta$ corresponding to
$\beta=0.1$, 0.5, 1.0 (thick curve) and 10.   
\label{fig5}} 
\end{figure}

Figure \ref{fig5} clearly shows at $t=82$~s the development of a pressure
enhancement and depletion, respectively, at the right and left
interfaces of the tube  with respect to the field free gas. This
essentially creates a pressure dipole which in turn generates a vortex
motion with upflow and downflow motions on the right and left sides,
respectively.  The top of the vortex motion is in the opposite
direction to that at the base. From Figure \ref{fig4} we find that, as time
proceeds, the vortex grows in size and also moves upward. This motion
carries the field lines, pushing them slightly counterclockwise in the
lower regions of atmosphere.  

As the vortex extends upwards, there is a hint of a shock forming at
its upper edge, as can be seen in Figure \ref{fig4}(c).  Once the transverse
driving motion stops at the base, the vortex motions diminish and the
flow is now guided along the field lines.  This is evident in Figure
\ref{fig4}(d) which shows at $t=136$~s negligible flows near the base of the
flux tube and large flows almost aligned with the magnetic field in
the upper atmosphere. The large upflows generate a shock which can be
discerned at a height of about 900 km along with a temperature
enhancement greater than 400 K.

\begin{figure} 
\epsscale{1.2} \hspace*{-0.7in}
\plotone{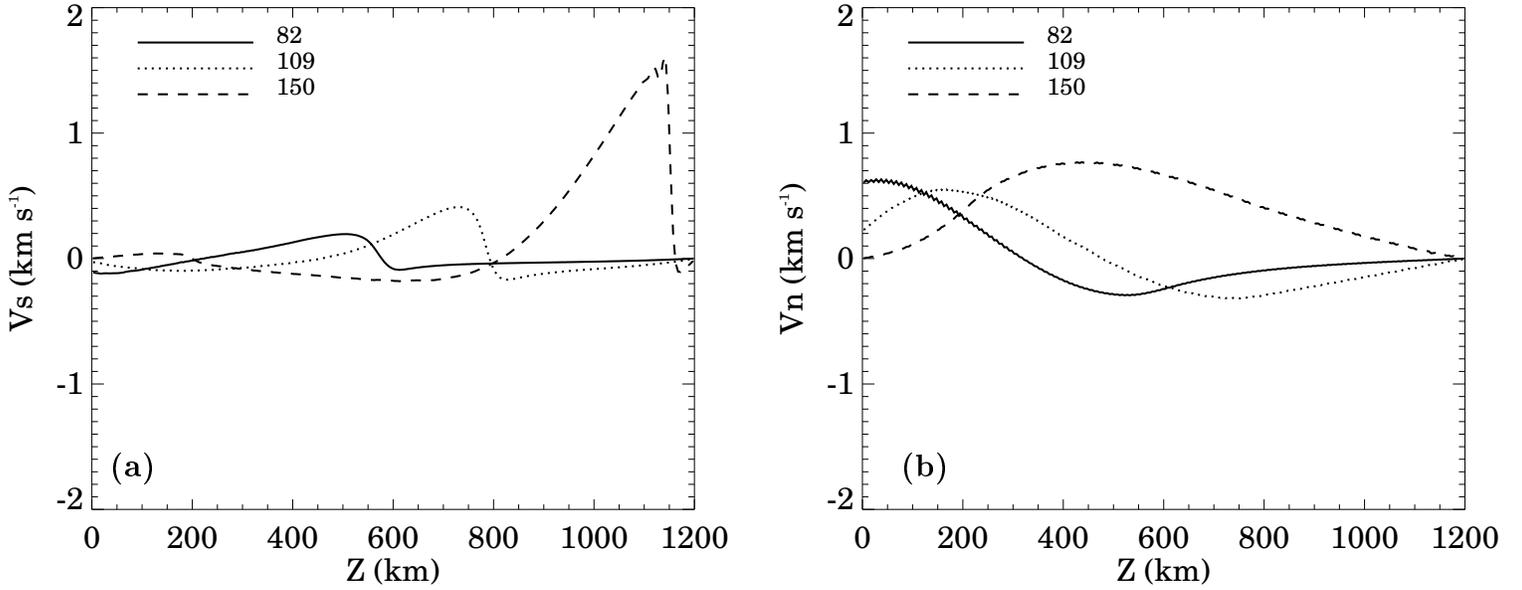}
\caption{
Variation of the velocity components (a) along the magnetic field $v_s$   
and (b) normal to the magnetic field $v_n$   
at different $t=82$~s (solid curve), 109~s (dotted curve) and
150~s (dashed curve) on a field line characterized by $f=0.6$
in a network element 
due to a horizontal motion at the lower boundary
with an amplitude of 750 m s$^{-1}$ and a period $P=240$~s.
\label{fig6}} 
\end{figure}

Figures \ref{fig6}(a) and (b) denote the $z$ variation of $v_s$  and $v_n$, the
field-aligned and normal velocity components respectively at the 
times of 82, 109 and 150~s along a specific field line. We
choose a field line parameterized by $f=0.6$, which is just 
to the right of the flux tube axis.  Let us first consider the
behavior of the field-aligned flow, shown in Figure \ref{fig6}(a). As time
proceeds, the vertical component of the velocity pulse increases with height in
magnitude on account of the density stratification and it steepens on
account of nonlinear effects as it propagates upward.  At $t=150$~s
the pulse resembles a shock.  The presence of a wake behind the shock
should also be noted, which is a generic feature associated with the
propagation of a pulse in a vertically stratified atmosphere (e.g.
Hasan et al.  2003).  Figure \ref{fig6}(b) shows the $z$ variation of the normal
component of the flow. The curves
corresponding to $t=$~82 and 109~s show the normal component
profiles during the impulsive phase (i.e., during the interval of
transverse driving motion at the base is present).  The dashed
curve shows the profile at $t=150$~s, after the transverse motion at
the base has halted. This profile is broadened due to the fact that
the Alfv\'en speed, which is the characteristic propagation speed of
the transverse pulse, increases sharply with height. In the lower
regions of the atmosphere, where $\beta\approx 0.6$, $v_A\approx 10$~km s$^{-1}$,
whereas at a height of 600 km, it has increased to about
26~km~s$^{-1}$.

\subsection{Periodic excitation} 
\begin{figure} 
\epsscale{1.2}\hspace*{-0.7in}
\plotone{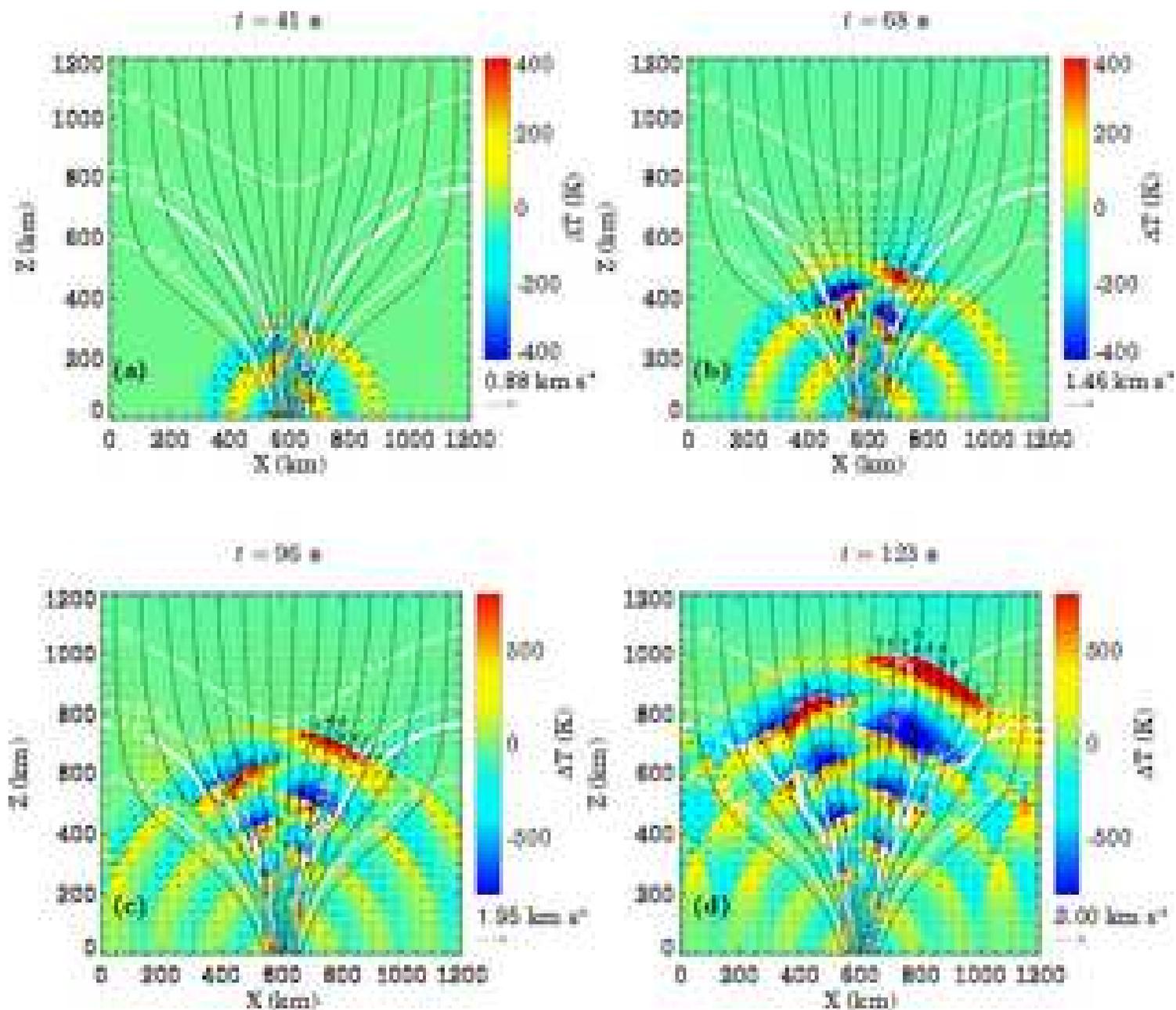} 
\caption  
{Flow pattern (arrows)  and the temperature perturbation (about the
initial state), $\Delta T$, at (a) 41~s, (b) 68~s, (c) 96~s and (d)
123~s in a network element due to a periodic horizontal motion at the
lower boundary with an amplitude of 750 m s$^{-1}$ and a wave
period of $P=24$~s.  The thin black curves are the magnetic field lines.
The white curves denote contours of constant $\beta$ 
corresponding to $\beta=0.1$, 0.5, 1.0 (thick curve) and 10.   
\label{fig7}} 
\end{figure}

We now consider the periodic excitation of waves due to transverse
driving of the lower boundary, with a period of 24~s (similar to
Rosenthal et al. and Bogdan et al.), though with a higher amplitude of
750~m s$^{-1}$.  Figures \ref{fig7}(a)-(d) show the wave pattern that
develops in this case.  As before, the colors are used to denote the
temperature perturbation with respect to the initial value at each
height and the white curves depict contours of constant $\beta$.  

The horizontal motion of the tube at the lower boundary is a source of
acoustic waves just outside the flux tube boundary in the ambient
medium, in which they propagate isotropically and near the interface as
`spherical' acoustic waves. The yellow and blue strips associated with
$\Delta T$, with an almost constant separation, clearly show the
acoustic waves propagating outwards at the sound speed from their
source near the bottom edge of the tube.  The wavelength of these
waves is approximately constant since the acoustic speed varies weakly
in the lower part of the computational domain.

\begin{figure} 
\epsscale{0.8}
\plotone{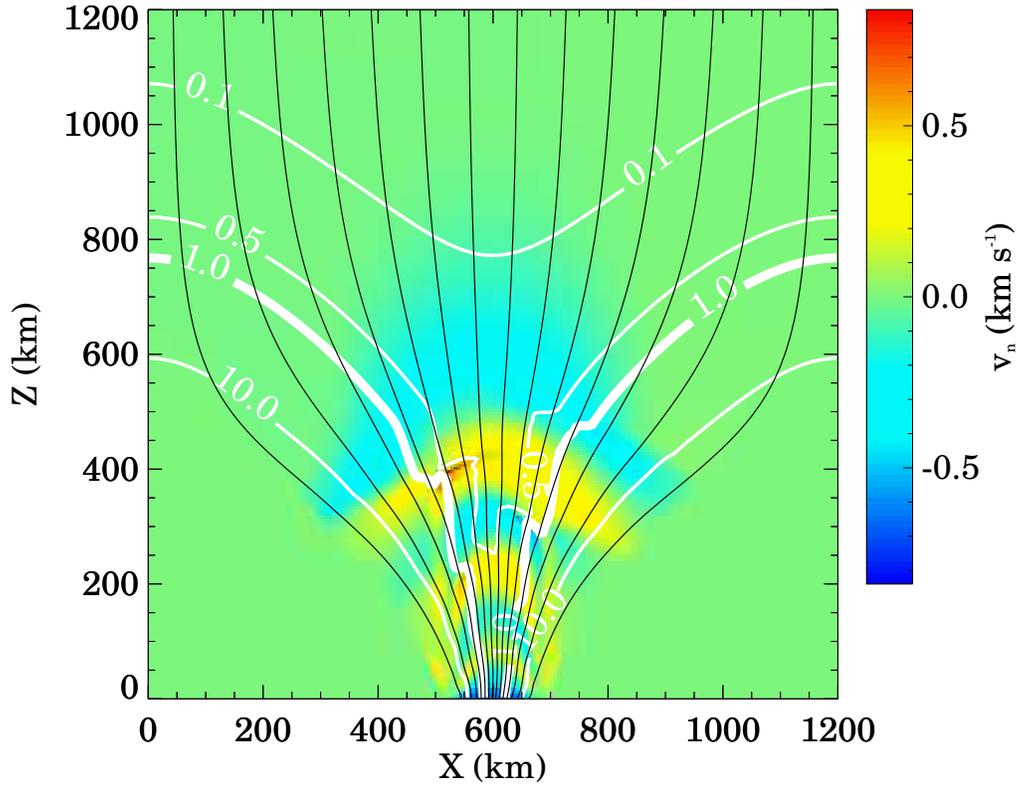}
\caption{
The normal velocity component $v_n$ 
at $t=68$~s
in a network element due to a horizontal motion at the lower boundary
with an amplitude of 750 m s$^{-1}$ and a period of $P=24$~s.
The thin black curves are
the magnetic field lines and the white
curves denote contours of constant $\beta$, corresponding to
$\beta=0.1$, 0.5, 1.0 (thick curve) and 10. 
\label{fig8}} 
\end{figure}

In the flux tube, the horizontal driving motions generate both fast
and slow modes. Close to the tube axis, the field is strong ($\beta <
1$) and the transverse motion generates a fast mode which propagates
upwards at the Alfv\'en speed. Its propagation can be clearly seen in
Figure \ref{fig8}, which depicts the variation of the normal velocity
component $v_n$ as a color contour plot. The fast wave front shows an
asymmetry in propagation since the Alfv\'en speed is largest on the
tube axis and decreases in the horizontal direction.  The increase in
separation in the vertical direction between the color peaks clearly
shows the increase in wavelength due to the increase in Alfv\'en speed
with $z$, which increases very rapidly with height from about 10 km
s$^{-1}$ at the base to about 100 km s$^{-1}$ at the top of our
computational box.

\begin{figure} 
\epsscale{0.8}
\plotone{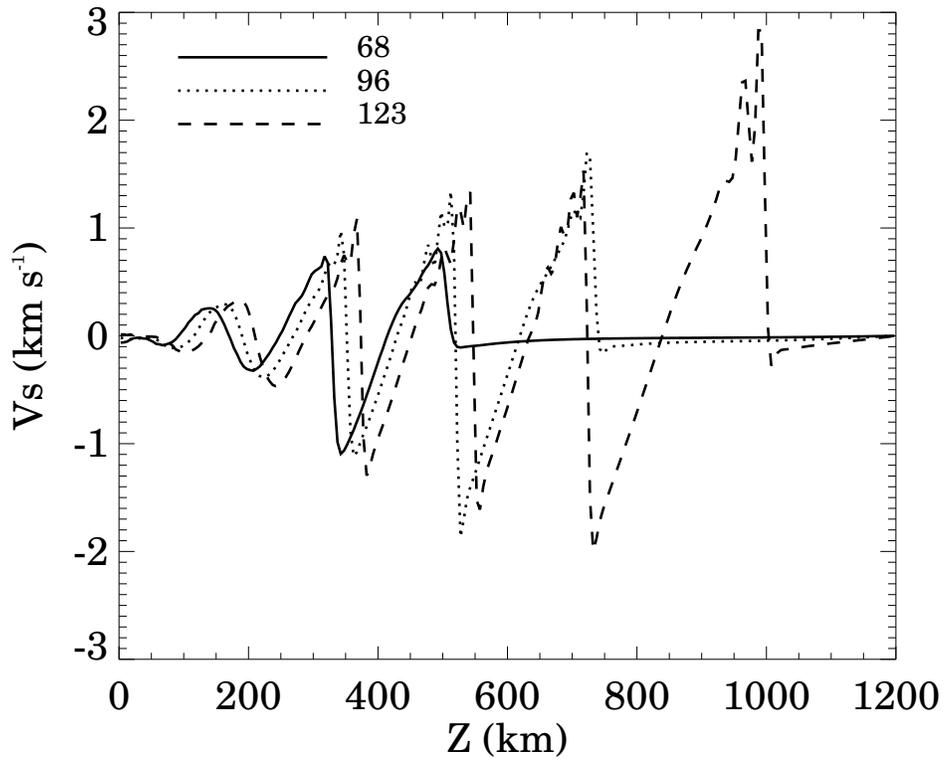}  
\caption
{
Variation of the velocity component along the magnetic field, $v_s$,   
at times $t=68$~s (solid curve), 96~s (dotted curve) and
123~s (dashed curve) on a field line characterized by $f=0.6$
in a network element 
due to  horizontal motion at the lower boundary
with an amplitude of 750 m s$^{-1}$ and a period of $P=240$~s.
\label{fig9}} 
\end{figure}

The horizontal motions in the central part of the flux tube produce
compressions and decompressions of the gas at the edge of the tube,
where the magnetic field is weaker ($\beta \ga 1$). The resulting
pressure variations are periodic in height and time, and are
$180^\circ$ out of phase on the two sides of the flux tube. The
associated pressure gradients drive periodic vertical flows that
propagate upward along the edge of the flux tube at the speed of the
transverse wave. These vertical flows are an integral part of the flux
tube wave, so the wave has both transverse and longitudinal character
in different parts of the tube. As the wave travels upward, it reaches
a height where the magnetic field at the edge of the tube becomes more
horizontal, and longitudinal motions are driven into the central part
of the flux tube where $\beta < 1$; this process is analogous to the
mode conversion discussed by Bogdan et al.\ (2003). In the low-$\beta$
region, the wave consists of two almost decoupled parts: a fast mode
that propagates in all directions, described in the preceding
paragraph (see Figure \ref{fig8}), and two slow modes that are guided
along the field lines. The slow modes on opposite sides of tube axis
are $180^\circ$ out of phase with each other, and their amplitudes
increase with height due to the stratification. Being mainly acoustic
in nature, they produce a compression and heating of the plasma in the
chromosphere. This can be discerned for instance in Figure
\ref{fig7}(d), where the wave front located at a height of about a
1000~km produces heating greater than 500 K. Figure \ref{fig9} shows
the variation of the field-aligned velocity component, $v_s$, on a
field line $f=0.6$ at three different times, shown by the different
curves.  One can see the amplitude increase and steepening of the wave
due to nonlinear effects as it propagates upward at the acoustic speed
of around $8$~km s$^{-1}$.  This clearly establishes  the nature of
this wave to be acoustic.  It is interesting to note the occurrence of
weak shocks at a Mach number of about 0.2. 

\section{Discussion and Summary} The present investigation is a
continuation of our earlier work on the dynamics of the magnetic
network on the Sun. In our previous studies we used a simplified
picture of magnetic elements in terms of a quasi one-dimensional
treatment based on the thin flux tube approximation. This approach was
useful in providing a qualitative picture of wave propagation and
nonlinear mode coupling in the lower regions of the atmosphere where
the thin flux tube approximation is applicable. However, as already
pointed out in Sect.\ 1, this approximation breaks down in the
chromosphere due to the expansion of the flux tube with height as well
as due to the merging of different tubes. We overcome these
limitations by using a two-dimensional treatment. In the first part of
this study we focus on a single flux tube and examine the consequences
when the lower boundary is perturbed by a transverse motion. Future
work will extend the present analysis to multiple tubes as well as
other types of excitation mechanisms.

The choice of using a transverse velocity perturbation at the lower
boundary was to some extent motivated by observations, particularly
those of Muller and Rodier (1992) and Muller et al.\ (1994) who
studied the footpoint motions of a large number of network bright
points, generally regarded as a proxy for magnetic elements.  We
considered the driving motions to occur at a fixed height in the
atmosphere. In reality, we expect that the displacement of the flux
tubes occurs in response to turbulent motions below the photospheric
base. These flows are absent or weak on the surface. Thus, the
pressure fluctuations in our model are a direct consequence of the
relative motion of the flux tube with respect to its local
surroundings. These play a key role in driving the vortex motions that
propagate high up in the flux tube.  It appears reasonable to expect
that such pressure fluctuations indeed occur in the solar network.

We chose two limiting forms for the time behavior of the perturbation
as an idealization for drivers that (a) generate discrete pulses well
separated in time and, (b) are periodic and continuous. The values of
the time constant for the pulse in the first case and the wave period
in the second case were chosen because of practical considerations
related to the limit of not letting the simulation  exceed a total
time where reflections from the top boundary become important.

We find that the transverse driving motions at the lower boundary lead
to strong pressure perturbations in the field-free medium at
the tube interface. This is an efficient process for generating
vortical motions, recognized earlier by Shibata (1983) in two-dimensional
simulations, who found that such motions can arise due to a pressure
perturbation applied at the base of a uniformly magnetized stratified
atmosphere ($\beta >1$). In our simulations the strong pressure
fluctuations are essentially localized in regions where $\beta$ is
greater than unity.  The observational signature associated with vortex
formation would be the prediction of simultaneous upward and downward
motions on opposite sides of flux tubes.  Under what conditions does a
vortex form? It appears that for wave periods that are sufficiently
long (compared to the Alfv\'en travel time) the dynamics is likely to
be dominated by vortical motions, but it is difficult at this stage to
say anything more definitive.

An important consequence of our calculations is that the interface
between the flux tube and the ambient medium is both a source of acoustic
waves in the ambient medium as well as fast and slow waves just inside
the flux tube. This can have interesting consequences: the acoustic
waves that travel isotropically in the field free medium will impinge
on neighboring flux tubes in the network and excite waves in them.
On the other hand the modes generated in the high $\beta$ region of
the tube (at the interface) undergo mode conversion as they propagate
upward and enter the region where $\beta\approx 1$, as also found by
Rosenthal et al.\ (2002) and Bogdan et al.\ (2003). In the upper part of
the flux tube where the field is essentially vertical and $\beta\ll 1$,
the acoustic-like longitudinal motions steepen and form shocks
accompanied by heating, as also found out for instance by Hasan \&
Ulmschneider (2004) using a thin flux tube calculation. There is,
however, an important qualitative difference between the present
results and those of Zhugzhda et al.  1995, Hasan et al. (2003) and
Hasan \& Ulmschneider (2004). In the latter, longitudinal motions are
generated from transverse driving motions at the flux tube footpoints
as a consequence of nonlinear mode coupling that is most efficient
when the transverse and longitudinal mode speeds are comparable
(roughly for $\beta\approx 0.2)$. In the present work, the mode
conversion occurs essentially in the linear regime and is a
consequence of the spatial variation of $\beta$ in the tube.
We find that transverse driving motions with velocities
less than 1 km s$^{-1}$ are sufficient to produce strong shocks
in the chromosphere.

The present work is in the spirit of the investigations by Rosenthal
et al.\ (2002) and Bogdan et al.\ (2003) on wave propagation in
two-dimensional magnetic structures.  There are, however, qualitative
differences between the initial state and the form of the perturbation
used for driving the atmosphere in the two sets of calculations.  For
instance, the initial equilibrium  configuration adopted by the above
authors assumes a potential field that is typical of the large-scale
pattern connecting different network regions. Bogdan et al.\ (2003)
consider a unipolar magnetic structure which at the base is 2000 km
wide surrounded by smaller magnetized regions (about 750 km wide) of
opposite polarity (as shown in Figure 1 of their paper).  At larger
horizontal distance from the flux concentration, the field is
essentially horizontal though much weaker. Thus, in the above
configuration the entire atmosphere is magnetized.  On the other hand,
we consider a discrete unipolar vertical flux, with a transverse
dimension of about 200 km, in magnetostatic equilibrium (and
non-potential field). Our equilibrium model, possessing a sharp
interface between the tube and the ambient field-free medium, is
perhaps more representative of a flux element in the solar network. A
minor difference is that the atmosphere in our model is not isothermal
in the vertical direction but has a temperature increase with height
similar to that in the solar chromosphere.  

Secondly, in our model, the whole flux tube at the photosphere is
displaced and not just a small region inside it.  The driving motion
in the simulations of Bogdan et al.\ (2003) is confined to a region
about 400 km wide within the flux element. In our simulations, the
transverse motion of the entire magnetic element creates strong
pressure perturbations just outside the tube which in turn are
responsible for the vortical motions and the generation of acoustic
waves in the ambient medium as well as the strong longitudinal motions
that eventually produce shock heating in the upper atmosphere.  The
recognition that the flux tube interface is a source of acoustic waves
in the ambient medium and that these waves can penetrate back into the
tube at higher elevations is a new feature of the present work.
Thirdly, in addition to periodic driving, we also consider impulsive
motions at the lower boundary.  

We should point out some limitations of our current study. Our
analysis is based on a two-dimensional treatment in slab geometry,
assuming that fluid displacements are confined to the $x-z$ plane. The
waves examined by us are different from the kink and sausage modes
treated earlier by us using a simplified thin flux tube approximation.
Ideally, one would use a 3-D treatment which is presently beyond the
scope of this work. Our analysis also neglects the torsional
Alfv\'en wave, which in reality would couple to the fast and slow
modes. Even within the two-dimensional framework, we have had to
restrict the height range in our simulations to 1200 km, since the
Alfv\'en speed (that essentially controls the time step used in the
numerical scheme) increases very rapidly with $z$. 

In summary this paper is the first of a series of investigations on
the dynamics of the solar magnetic network based on a two-dimensional
treatment of the MHD equations. We have made a start by considering
processes occurring in a single flux tube which expands with height in
the photosphere and assumes a "wine glass" geometry in the
chromosphere. Flows and wave motions are generated in this
configuration by transverse motions at the base of the flux tube. For
impulsive driving we find the presence of vortical motions. An
interesting feature for both impulsive and periodic driving, is the
development of shock-like features in the upper atmosphere which can
be important in heating the chromosphere. We hope to extend the scope
of the present calculation in future work to include multiple flux
tubes as well as include a larger height extension in order to
estimate transport of energy in to the corona through this
mechanism.

\acknowledgments
SSH thankfully acknowledges support from the National Science 
Foundation, U.S.A. through grant number ATM-0207641.



\clearpage
\begin{deluxetable}{llllll}
\tablecolumns{6}
\tablewidth{0pc}
\tablecaption{Parameters on the Tube Axis \& Ambient Medium }
\tablehead{
\colhead{}    &  \multicolumn{2}{c}{Tube Axis} &   \colhead{}   &
\multicolumn{2}{c}{Ambient Medium} \\
\cline{2-3} \cline{5-6} \\
\multicolumn{1}{l}{Variable} & \multicolumn{1}{l}{Base} & \multicolumn{1}{l}{Top} &
        \colhead{} & \multicolumn{1}{l}{Base} & \multicolumn{1}{l}{Top}  } 
\startdata
Temperature & 4800 K & 8200 K & & 4800 K & 8200 K  \\
Density & 1.3 10$^{-7}$ g cm$^{-3}$ & 9.5 10$^{-12}$ g cm$^{-3}$ & & 
          4.0 10$^{-7}$ g cm$^{-3}$ & 2.9 10$^{-11}$ g cm$^{-3}$  \\
Pressure & 4.1 10$^{4}$ dyn cm$^{-2}$ & 8.3  dyn cm$^{-2}$ & & 
          1.2 10$^{5}$ dyn cm$^{-2}$ & 22  dyn cm$^{-2}$  \\
Sound speed & 7.1 km s$^{-1}$  & 9.3 km s$^{-1}$  &  
            & 7.1 km s$^{-1}$  & 9.3 km s$^{-1}$  \\ 
Alfv\'en speed & 11 km s$^{-1}$  & 93 km s$^{-1}$  & & \nodata & \nodata \\ 
Magnetic field & 1420 G  & 102 G & & \nodata & \nodata  \\
$\beta$ & 0.5 K & 0.02 & & \nodata & \nodata  \\
\enddata
\end{deluxetable}

\clearpage

\clearpage

\clearpage
\clearpage

\clearpage

\clearpage

\clearpage

\clearpage


\end{document}